\documentclass[twoside,12pt]{article}%
\usepackage{amsmath}
\usepackage{epsfig}%
\usepackage{amsfonts}%
\usepackage{amssymb}%
\usepackage{graphicx}
\def\Journal#1#2#3#4{{#1} {#2} (#4) #3 }

\def\NPA{{\em Nucl. Phys.} A}

\def\PLB{{\em Phys. Lett.} B}

\def\PRL{\em Phys. Rev. Lett.}

\def\PREP{\em Phys. Rep.}

\def\PRD{{\em Phys. Rev.} D}
\def\PRC{{\em Phys. Rev.} C}

\newcommand{\be}{\begin{equation}}
\newcommand{\ee}{\end{equation}}
\newcommand{\bea}{\begin{eqnarray}}
\newcommand{\eea}{\end{eqnarray}}

\begin{document}

\title{ Properties of hadrons in a chiral model with (axial-)vector mesons}
\author{Francesco Giacosa\\\emph{Institut f\"{u}r Theoretische Physik }\\\emph{Johann Wolfgang Goethe - Universit\"{a}t, Max von Laue--Str. 1}\\\emph{D-60438 Frankfurt, Germany}}
\maketitle

\begin{abstract}
Recent advances in the development of a chiral linear $\sigma$-model with
(axial-)vector mesons are presented. The model is based on the basic
requirements of global chiral symmetry and dilatation invariance. The role of
(axial-)vector states turns out to be crucial both in the meson and the baryon
sectors. First results at nonzero temperature and density are discussed.

\end{abstract}

\section{Introduction}

The description of the masses and the interaction of low-lying hadron
resonances is a central subject of high-energy physics \cite{amslerrev}. To
this end, a lot of effort has been put in the development of quantum field
theoretical effective hadronic Lagrangians: chiral perturbation theory
\cite{chpt}, its extension with vector mesons \cite{chptvm}, and linear
$\sigma$-model(s) \cite{geffen,ko,buballa,denis,stani} represent the most
outstanding examples.

In this work we concentrate on the theoretical advances of the latter class of
hadronic theories: namely, we focus on a linear $\sigma$-model with
(axial-)vector mesons, which aims to describe (almost all) hadrons up to 1.7
GeV, both in the vacuum and at nonzero temperature and density. Although the
linear $\sigma$-model with scalar and pseudoscalar mesons has been extensively
studied, its generalization with (axial-)vector mesons was up to now not
systematically investigated. Preliminary studies in this direction have been
performed for the case $N_{f}=2$ only in Ref. \cite{ko,buballa}, where $N_{f}$
is the number of quark flavors. Moreover, in Ref. \cite{ko} a different
theoretical principle based on the so-called local realization of chirally
symmetry was employed, and in Ref. \cite{buballa} only `half' of the fields
was taken into account (the scalar isotriplet $a_{0}$ meson and the
pseudoscalar meson $\eta$ were neglected).

The construction of the Lagrangian of the here considered linear $\sigma$
model follows two basic requirements stemming from the underlying QCD theory
\cite{denis,stani,dynrec}: (i) Global chiral symmetry $U(N_{f})_{L}\times
U(N_{f})_{R}.$ (ii) Dilatation invariance, with the exception of terms
stemming from the gauge sector (scale-anomaly and $U(1)_{A}$ anomaly, in
accord with QCD) and of terms which describe the non-zero values of the quark
masses (explicit breaking of the dilatation symmetry). All other terms are
thus described by dimensionless coupling constants.

The assignment of the fields of the model with the resonances listed in the
PDG \cite{pdg} is straightforward for the multiplets $J^{PC}=0^{-+},$
$J^{PC}=1^{--}$, and $J^{PC}=1^{++}.$ As usual, the scalar meson multiplet
$J^{PC}=0^{++}$ is problematic: the present $\sigma$-model shows indeed
results which are at odd with previous $\sigma$-model studies. Namely, the
preferred scenario is realized for scalar mesons between $1$-$2$ GeV. This
result is a consequence of the inclusion of (axial-)vector degrees of freedom,
which generate peculiar interference effects in the decay amplitudes. More in
general, the role of the (axial-)vector states is relevant in both the meson
and baryon sectors for a proper description of the phenomenology.

Some of the resonances of the multiplet $J^{PC}=1^{++}$ were interpreted in
Ref. \cite{refs} as dynamically generated states. However, as discussed in
Ref. \cite{dynrec}, unitarization procedures can regenerate preformed
quark-antiquark states which were formally integrated out in order to obtain
the low-energy effective Lagrangians of Refs. \cite{chpt} and \cite{chptvm}.
Thus, the interpretation of the states $J^{PC}=1^{++}$ as a quark-antiquark
multiplet is, in agreement with Ref. \cite{pdg}, upheld\footnote{The example
of tensor mesons is clarifying: these states can be also obtained via
`dynamical generation', but fulfill to a very good accuracy all the required
properties to be interpreted as a quark-antiquark meson nonet \cite{tensormio}%
.}.

The present proceeding paper is organized as follows: in Sec. 2 and 3 we
review the mesonic and baryonic sectors respectively, and in Sec. 4 we present
our conclusions and outlooks.

\section{Meson sector}

The basic ingredients of the linear $\sigma$-model in the meson sector are the
glueball/dilaton field $G$, the (pseudo)scalar multiplet $\Phi=\left(
S^{a}+iP^{a}\right)  t^{a}$ and the left-handed and right-handed vectorial
multiplets $L^{\mu}=(V^{a,\mu}+A^{a,\mu})\,t^{a}$ , $R^{\mu}=(V^{a,\mu
}-A^{a,\mu})\,t^{a}$ (the matrices $t^{a}$ are the generators of the group
$U(N_{f})$). The mesonic Lagrangian $\mathcal{L}_{mes}$ which fulfills the
criteria of (i) global chiral symmetry and (ii) dilatation invariance takes
the following form for a generic number of flavors $N_{f}$ \cite{denis,stani}:%

\begin{align}
\mathcal{L}_{mes} &  =\frac{1}{2}(\partial_{\mu}G)^{2}-V_{dil}(G)+\mathrm{Tr}%
\left[  (D^{\mu}\Phi)^{\dagger}(D_{\mu}\Phi)-aG^{2}\Phi^{\dag}\Phi-\lambda
_{2}\left(  \Phi^{\dag}\Phi\right)  ^{2}\right]  -\lambda_{1}(\mathrm{Tr}%
[\Phi^{\dag}\Phi])^{2}+\nonumber\\
+ &  c(\det\Phi^{\dag}-\det\Phi)^{2}\text{ }+\mathrm{Tr}[\hat{\varepsilon
}(\Phi^{\dag}+\Phi)]-\frac{1}{4}\mathrm{Tr}[(L^{\mu\nu})^{2}+(R^{\mu\nu}%
)^{2}]\nonumber\\
&  +\frac{b}{2}G^{2}\mathrm{Tr}[(L^{\mu})^{2}+(R^{\mu})^{2}]+\frac{1}%
{2}\mathrm{Tr}[\hat{\delta}(L^{\mu})^{2}+(R^{\mu})^{2}]-2ig_{2}\left(
\mathrm{Tr}[L_{\mu\nu}[L^{\mu},L^{\nu}]]+\mathrm{Tr}[R_{\mu\nu}[R^{\mu}%
,R^{\nu}]]\right)  \nonumber\\
&  +h_{2}\mathrm{Tr}\left[  \Phi^{\dag}L_{\mu}L^{\mu}\Phi+\Phi R_{\mu}R^{\mu
}\Phi\right]  +2h_{3}\mathrm{Tr}\left[  \Phi R_{\mu}\Phi^{\dag}L^{\mu}\right]
\text{ }+...\text{,}%
\end{align}
where $D^{\mu}\Phi=\partial^{\mu}\Phi-ig_{1}(L^{\mu}\Phi-\Phi R^{\mu})$ and
dots represent further terms which are either large-$N_{c}$ suppressed or
unimportant in the evaluation of decays and (on-shell) scattering lengths.
Following comments are in order:

(i) Besides the dilaton field $G$, the mesonic fields of the model are
quark-antiquark fields. This can be easily seen by studying the so-called
large-$N_{c}$ limit \cite{thooft}: the masses are $N_{c}$-independent and the
widths scale as $N_{c}^{-1}$ \cite{denis}.

(ii) The dilaton potential reads $V_{dil}(G)=$ $=\frac{1}{4}\frac{m_{G}^{2}%
}{\Lambda_{G}^{2}}\left[  G^{4}\ln\left(  \frac{G}{\Lambda_{G}}\right)
-\frac{G^{4}}{4}\right]  $ \cite{schechter}. The parameter $\Lambda_{G}\sim
N_{c}\Lambda_{QCD}$ has dimension energy and sets the energy scale of the theory.

(iii) The $U(1)_{A}$ anomaly term is parametrized by the parameter $c$, which
has dimension [Energy]$^{4-2N_{f}}$.

(iv) In the (pseudo)scalar sector the explicit symmetry breaking of both
chiral and dilatation symmetries is encoded in the matrix $\hat{\varepsilon
}\propto diag\{m_{u},m_{d},m_{s},...\}$, where the entries are the bare quark
masses. Similarly, in the (axial-)vector sector the analogous diagonal matrix
$\hat{\delta}$ has been introduced.

(v) Chiral symmetry breaking takes place when the parameter $a$ is negative.
In fact, upon the condensation of the field $G=G_{0},$ the `wrong' mesonic
mass sign is realized for $aG_{0}^{2}<0$.

(vi) The calculations are performed at tree-level. The inclusion of loops is a
task for the future, but only slight changes are expected \cite{lupo}. Here we
mostly concentrate on decay widths; it is in this respect interesting to
notice that the latter do not correspond to a exponential law as function of
time because large variations take place for hadrons \cite{nonexp}.

Once the shifts of the scalar fields $G\rightarrow G_{0}+G$ and $\Phi
\rightarrow diag\{\sqrt{2}\sigma_{N},\sqrt{2}\sigma_{N},...\}+\Phi$, where the
first term is a diagonal matrix with the quark-antiquark condensates, and
necessary redefinitions of the pseudoscalar and axial-vector fields have been
performed, the explicit calculations of physical processes are lengthy but straightforward.

The case $N_{f}=2$ with frozen glueball $(m_{G}\rightarrow\infty)$ has been
studied in Ref. \cite{denis}. It has been shown that the inclusion of
(axial-)vector mesons has a strong influence on the overall phenomenology. For
instance, the width of the scalar meson $\sigma$ (the chiral partner of the
pion) decreases substantially w.r.t. the case without (axial-)vector mesons:
for this reason, the identification of this field with the resonance
$f_{0}(600)$ is not favoured, because the theoretically evaluated width is
smaller than $200$\ MeV, while the experimental one is larger than $400$ MeV.
On the contrary, the identification of the $\sigma$ field with the resonance
$f_{0}(1370)$ turns out to be in agreement with the experimental results. The
description of the (axial-)vector resonances is also in agreement with the
experiments reported in Ref. \cite{pdg}.

In Ref. \cite{stani} the glueball with a bare mass of about $m_{G}\sim1.5$
GeV, in agreement with the lattice calculation of Ref. \cite{latticeglueball},
has been studied for the first time in a chiral model with (axial-)vector
mesons. (For a compilation of other approaches see Refs. \cite{varieglue} and
refs. therein). The state $f_{0}(1500)$ results as the predominantly ($75\%$)
glueball state, and the rest of the phenomenology is only slightly affected
w.r.t. the previous case, in which $m_{G}\rightarrow\infty$. Moreover, also
the gluon condensate can be evaluated and turns out to be in agreement with
lattice results.

Future works in the meson sector consist in the study of the cases $N_{f}=3$
and $N_{f}=4$, respectively. Preliminary results of the former case were
already presented in conference proceedings \cite{nf3} and a systematic study
will be concluded soon. Interestingly, when increasing $N_{f}$ of one unit,
only two additional parameters, both related to the included bare quark mass,
are necessary. Thus, enlarging the model is straightforward (although it
implies lengthy calculations) and allows for more stringent checks of the model.

Finally, being the scalar states above $1$ GeV, a natural question is the
assignment of the existing resonances below $1$ GeV. According to the Jaffe's
interpretation \cite{jaffeorig} these resonances can be interpreted as a
scalar nonet of tetraquark states, see Refs. \cite{varietq} and refs. therein.
The coupling of tetraquark fields to the chiral model is possible by following
the prescription of Ref. \cite{tqmio}. The study of this scenario and the
arising mixing patterns represents an outlook. Interestingly, in the
simplified case $N_{f}=2$ there is only one scalar tetraquark field $\chi$,
which can be identified with the resonance $f_{0}(600)$; its coupling to the
chiral model has been discussed also at nonzero $T$ in Ref. \cite{achim},
where it has been shown that it can substantially influence the properties of
the chiral phase transition. This study shows that the correct interpretation
of scalar resonances is mandatory not only for a correct understanding of
spectroscopy in vacuum, but also toward a proper description of hadrons at
nonzero temperature.

\section{Baryon sector}

The inclusion of the nucleons and their chiral partner into the model was
performed in Ref. \cite{susanna}. The aim is the description of vacuum
phenomenology (such as pion-nucleon scattering lengths and decays), the origin
of the nucleon mass, the properties of nuclear matter, and the chiral phase
transition at nonzero density.

In the so-called mirror assignments \cite{detar} one starts with two baryonic
fields $\Psi_{1}$ and $\Psi_{2}$ transforming under chiral transformation as
\begin{equation}
\Psi_{1R}\longrightarrow U_{R}\Psi_{1R},\quad\Psi_{1L}\longrightarrow
U_{L}\Psi_{1L}\;,\Psi_{2R}\longrightarrow U_{L}\Psi_{2R},\quad\Psi
_{2L}\longrightarrow U_{R}\Psi_{2L}\quad.\label{mirror}%
\end{equation}
While $\Psi_{1}$ transforms as usual, $\Psi_{2}$ transforms in a `mirror way'.
Due to this characteristic it is possible to write down the following
Lagrangian in the baryonic sector for $N_{f}=2$:
\begin{align}
\mathcal{L}_{bar}  & =\overline{\Psi}_{1L}i\gamma_{\mu}D_{1L}^{\mu}\Psi
_{1L}+\overline{\Psi}_{1R}i\gamma_{\mu}D_{1R}^{\mu}\Psi_{1R}+\overline{\Psi
}_{2L}i\gamma_{\mu}D_{2R}^{\mu}\Psi_{2L}+\overline{\Psi}_{2R}i\gamma_{\mu
}D_{2L}^{\mu}\Psi_{2R}\nonumber\\
& -\widehat{g}_{1}\left(  \overline{\Psi}_{1L}\Phi\Psi_{1R}\ +\overline{\Psi
}_{1R}\Phi^{\dagger}\Psi_{1L}\right)  -\widehat{g}_{2}\left(  \overline{\Psi
}_{2L}\Phi^{\dagger}\Psi_{2R}\ +\overline{\Psi}_{2R}\Phi\Psi_{2L}\right)
+\mathcal{L}_{\mathrm{mass}}\text{ ,}\label{lbar}%
\end{align}
whereas $D_{iR(L)}^{\mu}=\partial^{\mu}-ic_{i}R(L)^{\mu}$ with $i=1,2$ and the
chirally invariant mass term $\mathcal{L}_{\mathrm{mass}}$ reads
\begin{equation}
\mathcal{L}_{\mathrm{mass}}=-\left(  \alpha\chi+\beta G\right)  (\overline
{\Psi}_{1L}\Psi_{2R}-\overline{\Psi}_{1R}\Psi_{2L}-\overline{\Psi}_{2L}%
\Psi_{1R}+\overline{\Psi}_{2R}\Psi_{1L})\;\text{.}\label{m0}%
\end{equation}
The chirally invariant fields $\chi$ and $G$ represent the scalar tetraquark
and glueball, respectively. (Note, the term $\mathcal{L}_{\mathrm{mass}}$
would not be possible if the field $\Psi_{2}$ would transform as $\Psi_{1}$.)
All the introduced coupling constants are dimensionless, as required by the
requirement of dilatation invariance.

Upon generation of the tetraquark and gluon condensates $\chi_{0}$ and $G_{0}%
$, a nonvanishing chiral mass emerges%
\begin{equation}
m_{0}=\alpha\chi_{0}+\beta G_{0}\text{ ,}%
\end{equation}
which represents the mass contribution to the nucleon which does \emph{not}
stem from the chiral (quark-antiquark) condensate $\sigma_{N}=\phi$. The
nucleon and its partner arise upon diagonalization of $\Psi_{1}$ and $\Psi
_{2}$:
\begin{equation}
\left(
\begin{array}
[c]{c}%
N\\
N^{\ast}%
\end{array}
\right)  =\frac{1}{\sqrt{2\cosh\delta}}\left(
\begin{array}
[c]{cc}%
e^{\delta/2} & \gamma_{5}e^{-\delta/2}\\
\gamma_{5}e^{-\delta/2} & -e^{\delta/2}%
\end{array}
\right)  \left(
\begin{array}
[c]{c}%
\Psi_{1}\\
\Psi_{2}%
\end{array}
\right)  \text{ , }\cosh\delta=\frac{m_{N}+m_{N^{\ast}}}{2m_{0}}\text{
,}\label{mix}%
\end{equation}
whereas the masses of $N$ and $N^{\ast}$ are a combination of $m_{0}$ and the
chiral condensate $\phi$:
\begin{equation}
m_{N,N^{\ast}}=\sqrt{m_{0}^{2}+\left(  \frac{\widehat{g}_{1}+\widehat{g}_{2}%
}{4}\right)  ^{2}\phi^{2}}\pm\frac{(\widehat{g}_{1}-\widehat{g}_{2})\phi}%
{4}\text{ .}\label{nuclmass}%
\end{equation}

The study of the vacuum's properties of this model has been presented in Ref.
\cite{susanna}. Upon a fitting to known experimental and lattice values, the
value of $m_{0}$ turns out to be non-negligible: $m_{0}=460\pm136$ MeV. As a
further step the pion-nucleon scattering lengths can be calculated and are in
agreement with the experimental data. (In particular, one has $a_{0}%
^{(-)}=\left(  6.04\pm0.63\right)  10^{-4}$ MeV$^{-1}$, which should be
compared with $a_{0}^{(-)\exp}=\left(  6.4\pm0.1\right)  10^{-4}$ MeV$^{-1}$.
The isospin-even scattering length $a_{0}^{(+)}$ depends on the scalar meson
sector and is therefore more subtle, see details in Ref. \cite{susanna}.) It
is in this respect important to stress that (i) (axial-)vector mesons play a
non-negligible role for the pion-nucleon scattering lengths and (ii) the use
of the naive assignment, which is equivalent to the case without the chiral
partner of the nucleon, delivers unacceptable result for the scattering lengths.

The extension to nonzero density has been investigated in Ref. \cite{susagiu},
in which the parameter $\beta$ in Eq. (\ref{m0}) has been set to zero. It is
thus assumed that the tetraquark state dominates: an important outcome of Eq.
(\ref{m0}) is the generation of a tetraquark-exchange term for the
nucleon-nucleon interaction, which intuitively corresponds to the exchange of
diquarks between nucleons. The corresponding resonance is the lightest scalar
state $f_{0}(600)$. As shown in Ref. \cite{susagiu} it is possible to describe
nuclear matter saturation for the very same value of $m_{0}$ obtained in the
vacuum. Moreover, the nuclear matter compressibility turns out to be in
agreement with the experimental value. For increasing density a first order
chiral phase transition takes place at $\rho\simeq2\rho_{0}$, at which both
quark-antiquark and tetraquark condensates drop to (almost) zero.
Interestingly, in the proposed approach the binding of nuclear energy is
possible in virtue of the existence of a light tetraquark field. An amusing
consequence is that nuclear matter seems to be a peculiarity of our $N_{c}=3$
world, which does not hold when increasing $N_{c}$ \cite{luca}.

\section{Conclusions and outlooks}

In this work recent results obtained with a linear $\sigma$-model with
(axial-)vector mesons have been presented. The inclusion of (axial-)vector
degrees of freedom is a necessary step toward a proper description of both
vacuum and medium properties of hadrons.

Various outlooks of the described approach are planned: (i) The extension to
$N_{f}=3$ is an ongoing project with preliminary results presented in Ref.
\cite{nf3}. Only two additional parameters w.r.t. the $N_{f}=2$ case are
needed. (ii) Inclusion of a light tetraquark nonet and evaluation of mixing
with the quark-antiquark states. (iii) The extension to $N_{f}=3$ allows to
included charmed mesons into the chiral model (for the first time with
(axial-)vector states). Also in this case only two additional parameters
w.r.t. the case $N_{f}=3$ are necessary. (iv) The pseudoscalar glueball with a
lattice predicted mass \cite{latticeglueball} of about $2.6$ GeV can be easily
incorporated into the model. Both outlooks (iii) and (iv) are interesting in
view of the planned Panda experiment \cite{panda} at the FAIR facility at
GSI/Darmstadt. (v) Evaluation of weak decays of the $\tau$-lepton into
hadrons. (vi) Calculation of nucleon-nucleon scattering and related dilepton
production, e.g. Ref. \cite{kk}. (vii) Extension of the study at nonzero $T$.
Preliminary works were performed with simplified versions of the models in
Refs. \cite{achim,stefan}, but a systematic study needs to be performed. Care
is needed in view of the discussion presented in Ref. \cite{achimnew}, where
it is shown that a modification of purely hadronic models is necessary in
order to be in agreement with basic properties of the large-$N_{c}$ limit.
(viii) Further studies of properties at nonzero density, with special
attention to the quarkyonic phase and to non-homogeneous condensates
\cite{quarkyonic}, are planned. The outlooks (vii) and (viii) can be combined
in order to achieve a complete picture of the chiral phase diagram of QCD.

\bigskip

\textbf{Acknowledgments: }the author thanks D. Parganlija, S. Janowski, S.
Gallas, G. Pagliara, L. Bonanno, P. Kovacs, A. Heinz, G. Wolf, and D. H.
Rischke for collaboration.

\end{document}